\documentclass[twoside,11pt]{article}
\usepackage{graphicx}
\usepackage{bm}
\usepackage{float}
\usepackage[round]{natbib}
\usepackage{booktabs}
\usepackage{lipsum}
\usepackage{cite}
\usepackage{appendix}
\usepackage[margin=1in]{geometry}
\usepackage[usenames]{color}
\usepackage{amsfonts, amssymb, amsmath, amsthm, multicol}
\usepackage[colorlinks=true, pdfstartview=FitV, linkcolor=blue,
citecolor=blue, urlcolor=blue]{hyperref}
\usepackage[usenames]{color}
\definecolor{Red}{rgb}{1,0.05,0}
\definecolor{Grn}{rgb}{0.1,0.7,0.1}
\definecolor{Blu}{rgb}{0.1,0.1,0.6}
\definecolor{Org}{rgb}{1,0.45,0}
\definecolor{Vio}{rgb}{0.6578,0,0.9478}
\definecolor{Mag}{rgb}{1,0.2,0.3}
\usepackage{authblk}
 \usepackage{booktabs}

\title{A Simple Boundary Condition Regularization
Strategy for Image Velocimetry Based Pressure Field
Reconstruction}

\author[1]{\textcolor{black}{Connor Pryce}}
\author[1]{\textcolor{black}{Lanyu Li}}
\author[2]{\textcolor{black}{Jared P. Whitehead}}
\author[1]{\textcolor{black}{Zhao Pan}\thanks{To whom correspondence may be addressed: zhao.pan@uwaterloo.ca}}
\affil[1]{University of Waterloo, Dept. of Mechanical and Mechatronics Engineering, Waterloo, ON, Canada}
\affil[2]{Brigham Young University, Mathematical Dept., Provo, UT, USA}

\date{\today}

\begin{document}
\maketitle \vspace{1cm}

\begin{abstract}
We propose a simple boundary condition regularization strategy to reduce error propagation in pressure field reconstruction from corrupted image velocimetry data. 
The core idea is to replace the canonical Neumann boundary conditions with Dirichlet ones obtained by integrating the tangential part of the pressure gradient along the boundaries. 
Rigorous analysis and numerical experiments justify the effectiveness of this regularization.
\end{abstract}

\section{Background and Statement of the Problem}




Solving the Pressure Poisson Equation (PPE) is one popular approach to pressure field reconstruction from image velocimetry data \citep{van2013piv}. Techniques like Particle Image Velocimetry (PIV) and Lagrangian Particle Tracking (LPT) provide a non-intrusive estimate of the pressure gradient $\nabla p = \bm g(\bm{u}) = -\bm{u}_t - (\bm{u} \cdot \nabla)\bm{u} + {\text{Re}}^{-1}\nabla^2 \bm{u}$ in the domain and on the boundary via the momentum equations, where $\bm{g}$ is a function of the velocity field $\bm{u}$, and $\text{Re}$ is the Reynolds number. 

Applying the divergence to $\nabla p$ yields the PPE:
\begin{equation}
    \label{eq: PPE ideal}
\nabla ^2 p  = f(\bm{u}) = -  \nabla \cdot  \left( \frac{\partial \bm{u} }{\partial t} + (\bm{u} \cdot \nabla )\bm{u} - \frac{1}{\text{Re}} \nabla^2 \bm{u} \right), 
\end{equation}
where $f$ is the data of the Poisson equation. 
To solve \eqref{eq: PPE ideal}, boundary conditions are required.
Commonly used boundary conditions are i)~Dirichlet boundary conditions
\begin{equation}
    \label{eq: PPE ideal DBC}
p  = h  \quad  \text{on} ~ \partial\Omega,  
\end{equation}
which define the value of the pressure on the boundaries and are often measured by pressure transducers and/or estimated using Bernoulli's principle in irrotational regions of the domain 
 \citep{de2012instantaneous}; as well as ii) Neumann boundary conditions  
\begin{equation}
    \label{eq: PPE ideal nNBC}
\bm{n} \cdot \nabla p   = g_n  \quad  \text{on} ~ \partial\Omega,  
\end{equation}
which define the pressure gradient normal to the boundary ($\bm{n}$ is the outward pointing unit normal vector on the boundary $\partial \Omega$), and are often derived from the momentum equations. 

While the Poisson equation benefits from the well-posedness of elliptic equations, superior numerical stability, ease of implementation, and high computational efficiency, experimentation (\citet{charonko2010assessment,sperotto2022meshless}; \citet{zhang2022uncertainty}) has shown that pressure field reconstruction on a domain with long Neumann boundaries may suffer from excessive error propagation due to contaminated image velocimetry data. 
This observation is expected, and supported analytically for two reasons: i) locally, the error in the data near a Neumann boundary is amplified and `diffuses' further into the interior of the domain \citep{faiella2021error}; and ii) globally, long Neumann boundaries result in a larger Poincare constant for the Laplacian operator and thus the error in the data may be further intensified when compared to a domain with Dirichlet conditions of the same length. For these reasons, and despite being available everywhere, Neumann boundary conditions should be avoided if possible~\citep{Pan2016Error1}.
Alternatively, Dirichlet conditions are favourable for taming error propagation, but they are not always easily accessible in practice. 

This renders a fundamental dilemma that inhibits the performance of the PPE for pressure field reconstruction from noisy velocimetry data. 
In an attempt to break this bottleneck, one idea is to replace the Neumann boundary conditions with derived Dirichlet ones.
In this letter, we introduce a simple regularization strategy of this type to address the boundary condition dilemma, and provide a rigorous analysis and validation to show that this specific regularization can improve the quality of the pressure field reconstruction.

\section{Boundary Condition Regularization}
\label{sec: strategy}

In addition to the natural choice of the canonical Neumann boundary conditions (cNBC) in \eqref{eq: PPE ideal nNBC}, the momentum equations also provide the pressure gradient tangential to the boundary:
\begin{equation}
    \label{eq: PPE ideal NBC t}
\bm{\tau} \cdot \nabla p   = g_\tau  \quad  \text{on}~ \partial\Omega,  
\end{equation}
where ${g}_\tau$ is the pressure gradient projected on $\bm{\tau}$, the unit vector tangential to the boundary in the counter-clockwise direction.
Assigning a reference pressure $p_0$ at a location $\xi_0$ on the boundary and integrating \eqref{eq: PPE ideal NBC t} directly gives the value of the pressure which is a derived Dirichlet boundary condition (dDBC):
\begin{equation}
    \label{eq: PPE ideal DDBC}
p(\xi) = h_d = \int_{\xi_0}^{\xi} \bm{\tau} \cdot \nabla p dS  + p_0 = \int_{\xi_0}^{\xi} g_\tau  dS + p_0  \quad  \text{on} ~ \partial\Omega. 
\end{equation}
Using this dDBC to replace the canonical Neumann boundary conditions for the Poisson equation yields a Dirichlet problem. 

\section{Analysis and Error Estimate}
\label{sec: error analysis}
We consider a simple example to see how the regularization strategy proposed in Sect.~\ref{sec: strategy} can improve the performance of the pressure field reconstruction based on a PPE with only a single point Dirichlet condition provided on the boundary. 
The error-contaminated problem for a domain with cNBC can be described as 
\begin{subequations}
\label{eq: PPE contaminated nNBC}
    \begin{alignat}{2}
      \nabla^2 \tilde p &=  \tilde f \quad &&\text{in}~ \Omega \label{subeqn: PPE contaminated} \\
      \bm{n} \cdot \nabla \tilde p  &=  \tilde g_n \quad &&\text{on}~ \partial \Omega, 
      \label{subeqn: nNBC contaminated}
    \end{alignat}
  \end{subequations}
where $\tilde p = p + \epsilon_p$, $\tilde{f} = f + \epsilon_{f}$, and $\tilde{g}_n = g_n + \epsilon_{g_{n}}$ are the contaminated pressure, data, and cNBC, respectively; and  
$\epsilon_p$, $\epsilon_{f}$, and $\epsilon_{g_{n}}$ are the corresponding error. 
Comparing \eqref{eq: PPE contaminated nNBC} with \eqref{eq: PPE ideal nNBC} and \eqref{eq: PPE ideal} leads to a new Poisson equation with respect to the error in the pressure field:
\begin{subequations}
\label{eq: error PE contaminated nNBC}
    \begin{alignat}{2}
      \nabla^2 \epsilon_p &=  \epsilon_f \quad &&\text{in}~ \Omega \label{subeqn: error PE contaminated} \\
      \bm{n} \cdot \nabla \epsilon_p  &=  \epsilon_{g_n} \quad &&\text{on}~ \partial \Omega, 
      \label{subeqn: error nNBC contaminated}
    \end{alignat}
  \end{subequations}
For the error propagation problem of \eqref{eq: error PE contaminated nNBC}, if we measure $\epsilon_p$ with a space-averaged error ($\| \epsilon_p \|_{L^2{\Omega}} = \sqrt{\int\epsilon_p^2 d\Omega/|\Omega|}$), the techniques in \citet{Pan2016Error1} provide an estimate for the error in the pressure field which we present here for convenience:
\begin{equation}
    \label{eq: error bound nNBC}
    \| \epsilon_p \|_{L^2(\Omega)} \leq C_N    \| \epsilon_f \|_{L^2(\Omega)} +  D_N \| \epsilon_{g_n} \|_{L^2(\partial\Omega)}, 
\end{equation}
where $D_N = \sqrt{C_NC_{NB}{|\partial\Omega|}/{|\Omega|}}$ is the coefficient corresponding to the error contribution from the boundary.
$C_N$ and $C_{NB}$ are the Poincare constants for the problem, independent of the image velocimetry technique, solver implementation and spatial resolution. These constants are solely determined by the geometry (i.e., shape, size, and dimension) of the domain and the setup of the boundary conditions.
$|\partial\Omega|$, $|\Omega|$ represent the size (length, area, or volume, depending on the dimension of the problem) of the boundary and the domain. 

If we apply the boundary regularization strategy described in Sect.~\ref{sec: strategy}, the error contaminated dDBC and the corresponding error are:
\begin{subequations}
\label{eq: PPE contaminated dDBC}
    \begin{alignat}{2}
      \tilde p  &=  \tilde h_d \quad &&\text{on}~ \partial \Omega \label{subeqn: PPE contaminated dDBC} \\
      \epsilon_p  &=  \epsilon_{h_d} \quad &&\text{on}~ \partial \Omega, 
      \label{subeqn: error dDBC}
    \end{alignat}
  \end{subequations}
where $\epsilon_{h_d}$ is the error in the dDBC. 
Replacing \eqref{subeqn: nNBC contaminated} with \eqref{subeqn: PPE contaminated dDBC}, the Poisson problem \eqref{subeqn: PPE contaminated} is regularized, and the corresponding error propagation can be modeled by replacing \eqref{subeqn: error nNBC contaminated} with \eqref{subeqn: error dDBC} in \eqref{eq: error PE contaminated nNBC}. 
Similar techniques in \citet{Pan2016Error1} provide an error estimate of the recovered pressure for a Dirichlet problem:
\begin{equation}
    \label{eq: error bound dDBC 1}
    \| \epsilon_p \|_{L^2(\Omega)} \leq C_D    \| \epsilon_f \|_{L^2(\Omega)} +  \| \epsilon_{h_d} \|_{L^\infty(\partial\Omega)}, 
\end{equation}
where $C_D$ is the corresponding Poincare constant. 
Noting that  $|\epsilon_{h_d}| \leq  \int_{\xi_0}^{\xi} |\epsilon_{g_\tau}| dS \leq |\partial \Omega| \| \epsilon_{g_\tau} \|_{L^2(\partial\Omega)}$, 
by the Cauchy-Schwarz inequality,
$\| \epsilon_{h_d} \|_{L^\infty(\partial\Omega)}$ in \eqref{eq: error bound dDBC 1} can be estimated as
\begin{equation}
    \label{eq: replace infy norm for dDBC}
    \| \epsilon_{h_d} \|_{L^\infty(\partial\Omega)} \leq 
    | \partial \Omega | \| \epsilon_{g_\tau} \|_{L^2(\partial\Omega)}.
\end{equation}
Combining \eqref{eq: error bound dDBC 1} and  \eqref{eq: replace infy norm for dDBC}, we arrive at
\begin{equation}
    \label{eq: error bound dDBC 2}
    \| \epsilon_p \|_{L^2(\Omega)} \leq C_D    \| \epsilon_f \|_{L^2(\Omega)} + | \partial \Omega | \| \epsilon_{g_\tau} \|_{L^2(\partial\Omega)},
\end{equation}
which is an estimate of the error in the reconstructed pressure field when the regularization described in \eqref{eq: PPE ideal DDBC} is practiced. 
For a domain with a given geometry, the Poincare constant ($C_D$ in \eqref{eq: error bound dDBC 2}) is almost always smaller
than that of the original Neumann problem ($C_N$ in \eqref{eq: error bound nNBC}. 
This suggests that the proposed regularization in \eqref{eq: PPE ideal DDBC} can improve the accuracy of the pressure reconstruction, especially for large domains.


Letting $\Omega = L\times L$ be a two-dimensional domain, a concrete example is presented to demonstrate the effectiveness of the proposed regularization. 
For the problem equipped with cNBC, \eqref{eq: error bound nNBC} becomes: 
\begin{equation}
    \label{eq: error bound nNBC LxL}
    \| \epsilon_p \|_{L^2(\Omega)} \leq \frac{L^2 }{\pi^2} \| \epsilon_f \|_{L^2(\Omega)}  +  \frac{4}{\pi^{3/2}} L \|\epsilon_{g_n} \|_{L^2(\partial\Omega)},  
\end{equation}
where $C_N = L^2/\pi^2$, $C_{NB} = 4L$, $|\Omega| = L^2$, and $|\partial \Omega| = 4L$. 
For the regularized problem on the same domain with dDBC, \eqref{eq: error bound dDBC 2} turns into
\begin{equation}
    \label{eq: error bound dDBC 2 LxL}
    \| \epsilon_p \|_{L^2(\Omega)} \leq \frac{L^2}{2\pi^2}  \| \epsilon_f \|_{L^2(\Omega)} + 4L\| \epsilon_{g_\tau} \|_{L^2(\partial\Omega)},
\end{equation}
where $C_D = L^2/2\pi^2$.
Comparing \eqref{eq: error bound nNBC LxL} against \eqref{eq: error bound dDBC 2 LxL}, it is obvious that the proposed regularization can produce a more accurate pressure reconstruction. 
For a square domain, $C_D$ is half of $C_N$, and an elongated rectangular domain reduces this factor again, further taming error propagation. 
This is achieved at the potential cost of slightly increasing the impact of the error in the pressure gradient on the boundary (the constant $4L/\pi^{3/2}$ in front of $\| \epsilon_{g_n} \|_{L^2(\partial\Omega)}$ in \eqref{eq: error bound nNBC LxL} is smaller than the constant $4L$ in front of $\|\epsilon_{g_\tau}  \|_{L^2(\partial\Omega)}$ in \eqref{eq: error bound dDBC 2 LxL}).
Thus, for experiments with high uncertainty in a large domain, the regularization is expected to be effective in reducing the error in the reconstructed pressure field.

\section{Curl-Free Regularization on a Closed Boundary}
\label{sec: curl free regulation}

If the regularization proposed in \eqref{eq: PPE ideal DDBC} is employed on a closed boundary, it may lead to a non-physical Dirichlet condition when the error on the boundary ($\epsilon_{g_\tau}$) is not curl-free. 
That is, after the integration indicated in \eqref{eq: PPE ideal DDBC}, the pressure value at $\xi=\xi_0$ may not be continuous ($p(\xi_0^+) \neq p(\xi_0^-)$).
If integrated along another direction starting from the same reference point and reference pressure, a different pressure along the boundary is expected.
This should not be the case since $\nabla p$ is curl-free (including the path along the boundary), and integration on $\nabla p$ should be path-independent. 
To resolve this issue, we construct a continuous function $p(\xi)$ on the boundary such that $\nabla p(\xi)$ is an interpolant or regression of $\tilde g_\tau$, subject to the constraint $p(\xi_0^+) = p(\xi_0^-) = p_0$. 


In 2D, one way to explicitly achieve such a Dirichlet boundary is by linear interpolation:
\begin{equation}
    \label{eq: pryce regularization}
   p(\xi) = \tilde{h}_d = \underbrace{(1 - \theta)\int_{\xi_0}^{\xi}  \bm{\tau} \cdot \nabla \tilde p  dS}_{\text{counter~clockwise~int.}} +  \underbrace{ \theta\int_{\xi_0 + |\partial\Omega|}^{\xi} \bm{\tau} \cdot \nabla \tilde p  dS}_{\text{clockwise~int.}} + p_0,
\end{equation}
where $\theta = (\xi-\xi_0)/|\partial \Omega|$, and $ \xi_0 \leq \xi \leq \xi_0+|\partial\Omega|$.
Equation \eqref{eq: pryce regularization} has an apparent physical meaning and can be thought of as a weighted average of integration of $\nabla \tilde{p}$ along two different directions, with $\theta$ being a weighting parameter.
For $\theta = 0~\text{or}~1$, the counter-clockwise or the clockwise integral starting from $\xi_0$ has not yet accumulated any error, and thus, $p(\xi_0^+) = p(\xi_0^-)=p_0$.
When $\theta \in (0,1)$, \eqref{eq: pryce regularization} is a linear interpolation between the two integrals, and it is  easy to verify that $p'(\xi) = \bm{\tau} \cdot \nabla \tilde{p} = \tilde{g}_\tau$.
This modified regularization guarantees that the resulting $\tilde{h}_d$ is continuous even if the contaminated pressure gradient on the boundary is not curl-free, and thus this regularization can be considered as a curl-free boundary correction. 
Note, \eqref{eq: pryce regularization} is one of the simplest solutions, but not necessarily the only one or the best one; nevertheless, its advantage will be demonstrated in Sect.~\ref{sec: validation}.           

\section{Numerical Experiments and Validation}
\label{sec: validation}

\begin{figure}
    \centering
    \includegraphics[width=0.6\columnwidth]{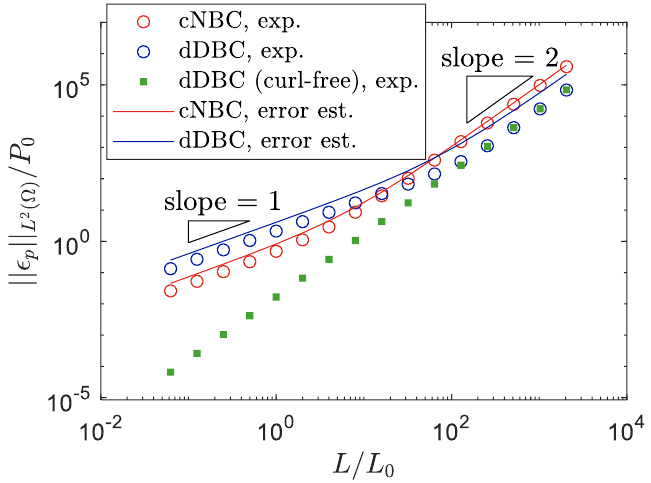}
    \caption{Error in the reconstructed pressure field ($||\epsilon_p||_{L^2(\Omega)}/P_0$) scaling with the length scale of the domain ($L/L_0$). Symbols represent the results from numerical experiments; the red and blue solid lines illustrate the error estimates indicated in \eqref{eq: error bound nNBC LxL} and \eqref{eq: error bound dDBC 2 LxL}, respectively.}
    \label{fig: scaling law}
\end{figure}

To validate the proposed boundary regulation strategies proposed in \eqref{eq: PPE ideal DDBC} and \eqref{eq: pryce regularization}, a synthetic flow based on the Taylor vortex contaminated by artificial error is considered numerically. 
The pressure field of the Taylor vortex is 
\begin{equation}
    \label{taylor vortex}
    p = -\frac{\rho H^2}{64 \pi^2 \nu t^3} \exp \left(-\frac{r^2}{2 \nu t}\right),
\end{equation}
where, $H$ represents the angular moment of the vortex, $\nu$ the kinematic viscosity of the fluid, $\rho$ the density of the fluid, $t$ the time, and $r$ the distance from the origin in polar coordinates. 
We choose the parameters so that the characteristic length scale of the vortex is $L_0 = \sqrt{2 \nu t} = 1$ and the leading coefficient $\frac{\rho H^2}{64 \pi^2 \nu t^3}$ as well as the amplitude of the characteristic pressure for this flow is unity ($P_0= 1$). 
In this numerical experiment, the ground truth of the pressure field, the pressure gradients (i.e., $g_n$ and $g_{\tau}$) and the pressure Laplacian $f$ are generated from \eqref{taylor vortex} on an $L \times L$ square domain. 
Bias error (i.e., $\epsilon_{g_n} = 1$, $\epsilon_{g_{\tau}} = 1$ and $\epsilon_f = -1$) were added to the true solution to generate the synthetic data.
This constant bias represents a global error with dominating low-frequency components and is typically challenging \citep{faiella2021error}.  
Using this synthetic data, the PPE was solved with the commonly used cNBC, the proposed dDBC, and the improved curl-free dDBC while varying the size of the domain. 
The reconstructed pressure was then compared to the ground truth to evaluate the error associated with each numerical experiment and is illustrated in Fig.~\ref{fig: scaling law}.


As seen in Fig.~\ref{fig: scaling law}, the results from the numerical experiments match the error estimates derived in Sect.~\ref{sec: error analysis} well.
For the cNBC setup, the error estimate in \eqref{eq: error bound nNBC LxL} is relatively sharp, while the upper bound in \eqref{eq: error bound dDBC 2 LxL} is slightly more conservative for the dDBC setup, meaning that the actual performance of the pressure solver with dDBC regularization is even better than expected. 
Both error estimates reveal the scaling behavior of the error in the pressure with the length scale of the domain $L$. 
For the two-dimensional cases, when the domain is large, the error in the domain ($\epsilon_f$) dominates and $\| \epsilon_p\|_{L^2(\Omega)} \sim L^2$; while for a small domain, error on the boundary ($\epsilon_{g_n}$ or $\epsilon_{g_\tau}$) dominates and $\| \epsilon_p\|_{L^2(\Omega)} \sim L^1$. 
The curl-free dDBC proposed in Sect.~\ref{sec: curl free regulation} also performs as expected, resulting in even less error than the two aforementioned methods, especially when the domain is small. 
This test showcases the power of these simple regularization strategies and validates the error estimates outlined in Sect.~\ref{sec: error analysis}.

\begin{figure}
    \centering
    \includegraphics[width=1\columnwidth]{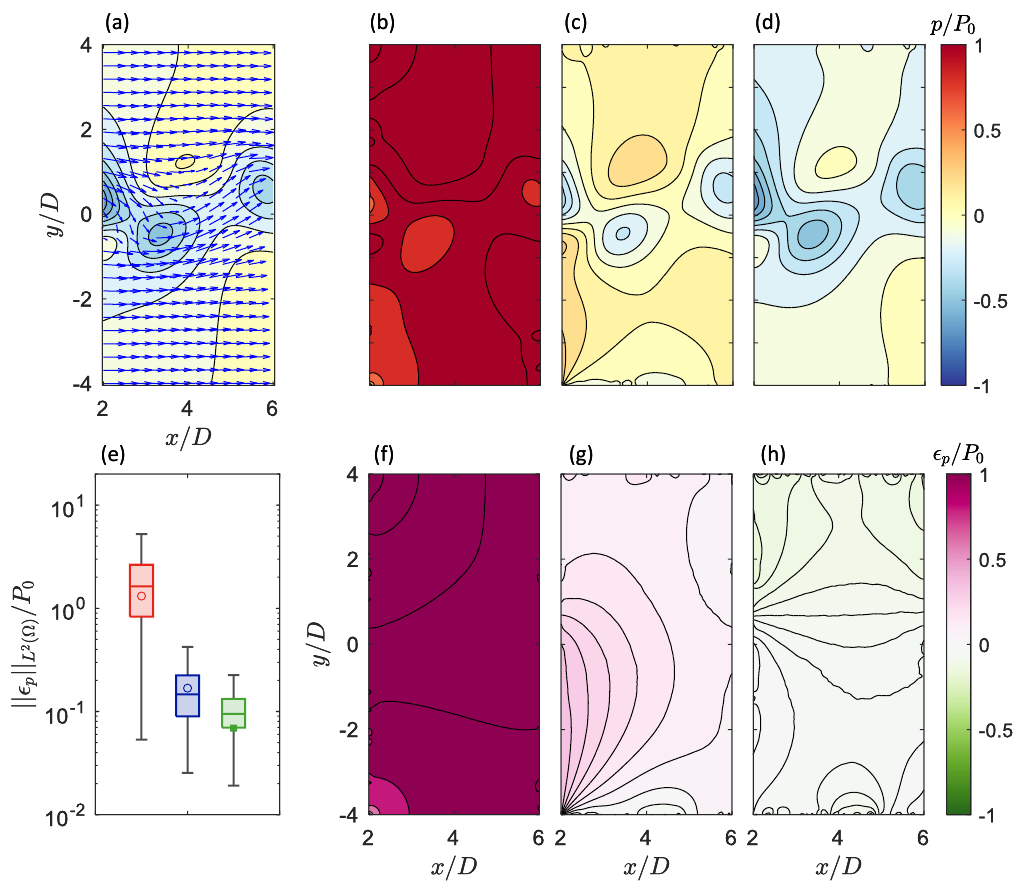}
    \caption{Typical results and statistical tests for pressure reconstruction with and without boundary regularization. (a) quiver plot of the velocity field overlaid on the pressure field ground truth; (b--d) reconstructed pressure field with cNBC, dDBC, and curl-free dDBC, respectively; (f--h) error in the reconstructed pressure field by comparing (b--d) with (a). (e) box plot of error in the pressure reconstruction from 500 independent tests with the red, blue, and green boxes corresponding to the statistics of the error for cNBC, dDBC, and curl-free dDBC cases, respectively. Horizontal bars in the middle of the boxes show the median while the upper and lower edges of the box indicate the 25 and 75 percentiles. The upper and lower whiskers bound the 95\% confidence intervals of the error while the symbols within the boxes mark where the corresponding error shown in (f--h) lie within the data. 
    }
    \label{fig: cylinder flow demo}
\end{figure}

The proposed boundary regularization strategies are further tested using synthetic data of a wake behind a cylinder with $Re = 100$, based on free stream velocity $U_\infty = 1$.
A cropped and down-sampled high-fidelity simulation of the wake flow was used as the ground truth of the velocity and pressure fields. 
The cylinder of diameter $D$ was centered at $(x/D,y/D) = (0,0)$, and the cropped domain is $(x/D,y/D) \in [2,6] \times [-4,4]$.
The non-dimensional spatial and temporal resolution of the down-sampled data is $dx/D = dy/D = 0.0625$ and $dtD/U_\infty =0.1$ respectively. 
Point-wise Gaussian noise with a variance equal to $1\%$ of the magnitude of the free stream velocity was added to the ground truth velocity to generate the synthetic data. 
We reconstruct the pressure field using a second-order finite difference Poisson solver
using the artificially contaminated velocity data and one point reference pressure at the bottom left corner of the domain $(x/D,y/D)=(2,-4)$ with cNBC, dDBC, and curl-free dDBC. 
The reconstructed pressure and the error are normalized by $P_0=\frac{1}{2}\rho U_\infty^2$.
500 independent tests were carried out and depicted in Fig.~\ref{fig: cylinder flow demo}.



The statistical tests in Fig.~\ref{fig: cylinder flow demo} show the performance of the three different boundary conditions with both the dDBCs performing much better than the cNBC as expected. 
In particular, the median error in the dDBC is reduced by roughly a factor of 10 when compared to the cNBC, and the curl-free dDBC reduces this error by an additional factor of $\sim$2. 
This is consistent with i) the theory that an elongated rectangular domain can reduce error propagation even further when compared to a square domain as indicated by \eqref{eq: error bound nNBC LxL} and \eqref{eq: error bound dDBC 2 LxL}; and ii) that the error estimate in \eqref{eq: error bound dDBC 2} is rather conservative. 
In addition to greatly reducing the expected error in pressure (i.e., the mean of $\| \epsilon_p \|_{L^2(\Omega)}/P_0$ in Fig.~\ref{fig: cylinder flow demo}(e) are 1.83, 0.17, and 0.11 for cNBC, dDBC, and curl-free dDBC case, respectively), the variance of $\| \epsilon_p \|_{L^2(\Omega)}/P_0$ is also reduced by the boundary regularization (e.g., the standard deviation of $\| \epsilon_p \|_{L^2(\Omega)}/P_0$ are 1.27, 0.10, and 0.05, cNBC, dDBC, and curl-free dDBC cases, respectively).
This means that the proposed boundary regularization can improve both the accuracy and precision of pressure reconstruction. 
The observation in this test can also be explained from the numerical perspective:  
The resulting discretized systems from the PPE (i.e., $\mathcal{L}p=f$, where $\mathcal{L}$ is the discretized Laplacian with associated boundary conditions) have 2-norm condition numbers of $1.1\times10^4$ and $4.6\times10^5$ with and without regularization, respectively, meaning that the regularization improves the numerical conditioning of the problem as well. 
The effectiveness of the boundary regularization can be directly visualized by Fig.~\ref{fig: cylinder flow demo}(b--d) or Fig.~\ref{fig: cylinder flow demo}(f--h), which is a typical case out of the 500 tests. 
Without boundary regularization, high bias in the reconstructed pressure ($\tilde{p}$) occurs due to the application of long Neumann boundaries, except for the bottom left corner where a transducer provides an accurate reference pressure (see Fig.~\ref{fig: cylinder flow demo}(b\&f)). 
From Fig.~\ref{fig: cylinder flow demo}(c\&g), it is obvious that replacing long cNBC with dDBC can effectively tame the error propagation; however,  error accumulating along the boundary due to the practice of the regularization of \eqref{eq: PPE ideal DDBC} in the counter-clockwise direction is propagated to the interior of the domain (i.e., the $\epsilon_p$ is higher along the left edge of the domain than that of the bottom edge). 
This artifact is resolved by the curl-free dDBC regularization (see Fig.~\ref{fig: cylinder flow demo}(d\&h)). This suggests that even the curl-free correction on the boundary alone typically results in further reduction in $\epsilon_p$ throughout the entire domain.

\section{Concluding Remarks and Perspectives}
This work portrays a simple yet overlooked idea to regularize the velocity-based pressure field reconstruction: replacing the canonical Neumann boundary conditions with the Dirichlet conditions derived from integrating the tangential part of the pressure gradient on the boundary. 
This simple practice can effectively improve the quality, in terms of accuracy and robustness, of the pressure field reconstruction from corrupted image velocimetry data, which is analytically proved and validated by numerical experiments.
The corresponding error estimates provided in this work are conservative i.e. pessimistic, but still dictate the generic dynamics of error propagation. 

The proposed regularization strategies are very simple, and presumably can make Dirichlet conditions available everywhere with low computational cost. This is attractive as the derived Dirichlet conditions provide a  `stronger' type~\citep{faiella2021error} of information on the boundary than cNBC alone.
This `additional' information on the boundary provides great flexibility and potential for more sophisticated algorithms to further improve the reconstruction. 
Our analysis is independent of the dimension of the domain (2D or 3D), and the continuous setting of the analysis indicates that the strategy works for structured or unstructured data based on PIV or LPT. 
Beyond establishing this family of regularization strategies, fulfilling the full potential of this seemingly odd, yet effective idea, will be left for future investigations.





\section{Acknowledgement}
We thank the discussion with Drs. Fernando Zigunov and Grady Wright.
This work is partially supported by the Natural Sciences and Engineering Research Council of Canada (NSERC) Discovery Grant (RGPIN-2020-04486), and the Undergraduate Research Assistantship (URA) program, University of Waterloo. J.P.W. was partially supported by NSF grant DMS-2206762.

\bibliographystyle{apalike}
\bibliography{PIV_Pressure_Lib}
\end{document}